\RequirePackage{fix-cm}
\documentclass[preprint,12pt]{elsarticle}

\usepackage[OT1]{fontenc}
\usepackage{amsmath,amssymb,bm}
\usepackage{graphicx}
\usepackage[hidelinks]{hyperref}
\usepackage{bbm}
\newcommand{\one}{\mathbbm{1}}
\newcommand{\R}{\check R}
\usepackage{color}

\journal{Nuclear Physics B}

\begin{document}
\begin{frontmatter}

\title{Exact fermionic decay length and Majorana localization in the sign-alternating kagom\'e Ising model}
\author[ansl]{Sh.~Khachatryan}
\author[ansl]{H.~Babujian}
\author[ansl]{A.~Sedrakyan}
\address[ansl]{Alikhanyan National Science Laboratory, Br.\ Alikhanian 2, Yerevan 0036, Armenia}

\begin{abstract}
We derive an exact fermionic decay length for the sign-alternating
kagom\'e Ising model at fixed transverse momentum $p=\pi/2$.
Writing $x=\tanh K_1$, $y=\tanh K_2$, and $z=\tanh K_3$ for the
three dimensionless couplings, the inverse length per lattice cell
is $\xi_{\rm f}^{-1}=|\ln(U/|V|)|$, where
$U=(1-x^2)(1+y^2z^2)$ and $V=(1+x^2)(y^2-z^2)$.
The length diverges at $U=|V|$, with the sign of $V$ selecting
one of two classical critical branches. In a simultaneous
strong- and weak-bond scaling, these branches become the
gap-closing conditions at momenta $0$ and $\pi$ of the
transverse-field Ising chain obtained from the row transfer matrix.
On the topological side, the classical decay length tends to
the localization length of a Majorana end mode of the
Jordan--Wigner fermion chain. We also prove that the regular
strong-bond Yang--Baxter family cannot be continued to the
finite-coupling interior within the fixed-basis ansatz with one
vanishing bond direction.
\end{abstract}


\end{frontmatter}

\section{Introduction}
\label{sec:introduction}

We study the decay of fermionic correlations in a two-dimensional
sign-alternating kagom\'e Ising model and its relation to Majorana
localization in the quantum chain obtained from the row transfer
matrix. The main result is an exact decay length in the momentum
channel containing two classical critical branches. Its scaling
limit gives the Majorana localization length on the topological side.

An Ising spin $\sigma_r=\pm1$ sits at each lattice site $r$. The
energy and partition function are
\begin{equation}
 \mathcal H_{\rm cl}=-\sum_{\langle rr'\rangle}J_{rr'}\sigma_r\sigma_{r'},
 \qquad
 \mathcal Z=\sum_{\{\sigma_r\}}\exp[-\mathcal H_{\rm cl}/(k_{\rm B}T)].
 \label{eq:classical-model}
\end{equation}
Here $T$ is the temperature, $k_{\rm B}$ is Boltzmann's constant,
and $\langle rr'\rangle$ runs over nearest-neighbor bonds.
Positive $J_{rr'}$ favors parallel spins. A face is frustrated when
the product of its bond signs is negative, because no spin
configuration can satisfy all bonds around that face. The
uniform kagom\'e antiferromagnet is a familiar example with
finite residual entropy \cite{kano}.

In each of the three bond directions $i=1,2,3$, the two
periodically repeated bond sets carry the dimensionless couplings
$K_i$ and $K'_i$, obtained from the bond energies by dividing by
$k_{\rm B}T$. We set $K'_i=-K_i$. Our earlier work developed the
free-fermion vertex representation for this model, identified
the shifted zeros of the partition-function polynomial, and
derived the quantum spin chain in the anisotropic limit \cite{kag}.
Here we use the momentum-space polynomial to obtain the exact
decay rate and relate it to the quantum chain's boundary modes.

\begin{figure}[tbp]
 \centering
 \includegraphics[width=.85\linewidth]{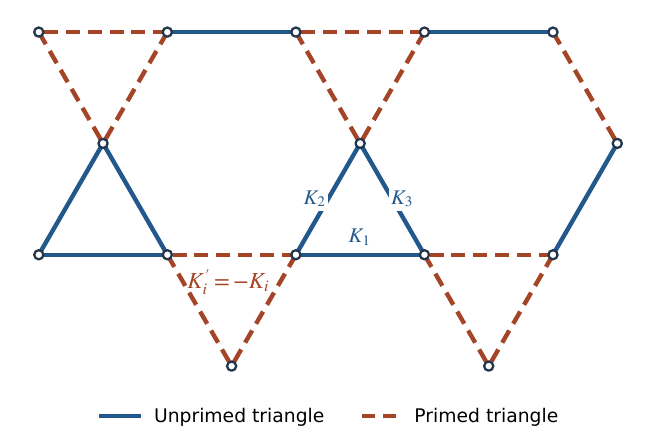}
 \caption{The two sets of nearest-neighbor bonds on the kagom\'e
 lattice. Solid bonds carry $K_i$ and dashed bonds carry
 $K_i'=-K_i$. The direction index $i$ is the same for parallel
 bonds. For the illustrative choice $K_i>0$ in all three
 directions, the solid triangles are ferromagnetic and the dashed
 triangles are frustrated. Open circles denote Ising sites.}
 \label{fig:kagome-bonds}
\end{figure}

Figure~\ref{fig:kagome-bonds} shows the two bond sets in the
sign-alternating model.

The momentum-space Kac--Ward determinant provides a general
framework for the spectral polynomials of periodic planar Ising
models \cite{Sedrakyan-1984,CimasoniDuminilCopin2013}.
At fixed transverse momentum $p=\pi/2$, we factor the kagom\'e
fermionic polynomial \cite{kag} and continue the other momentum
into the complex plane. The roots give the inverse decay length
$\xi_{\rm f}^{-1}=|\ln(U/|V|)|$ in Eq.~\eqref{eq:exact-decay}.
The two real-momentum zeros reached at $U=|V|$ become the gap
closings at $0$ and $\pi$ in the anisotropic quantum chain. The
same formula gives the spatial decay of a Majorana end mode
after taking the scaling limit on the topological side.

The vertex representation gives two distinct ways to study this
question. The free-fermion identity reduces the classical partition
function to a quadratic fermion problem. The Yang--Baxter equation {\color{blue} \cite{Baxter-book, Takhtajian-book}}
gives an algebraic construction of row transfer matrices that commute
as a spectral parameter varies \cite{kag,ks,kh}. We call the limit
$K_1\to+\infty$, $K'_1\to-\infty$ the \emph{strong-bond boundary}.
There the normalized checked matrix belongs to a regular
Yang--Baxter family. Regularity means that the checked matrix equals
the identity at a particular spectral value. At that value the row
transfer matrix is a lattice translation, and its logarithmic
derivative gives pure Ising exchange.

The transverse-field chain arises when $K_1$ grows while $K_2$ and
$K_3$ shrink at a prescribed rate. This brings the transverse field
and the exchange into the transfer-matrix expansion at the same
order. After a Jordan--Wigner transformation, the ordered spin
regime is a fermionic phase with a Majorana mode at each open end
\cite{pfeuty,kitaev}. The spin cohomology and fermionic bordism
classifications specify the role of the symmetries in this
one-dimensional system \cite{cgw,kttw,fidkowski}. Our second result is an obstruction to extending the boundary
Yang--Baxter family to finite couplings within the fixed-basis
ansatz with one vanishing bond direction. We obtain it by
differentiating the braid equation at coincident spectral
parameters.

 At zero temperature, the one-dimensional transverse-field
	Ising chain has a continuous quantum phase transition at $g=|J_q|$,
	where $g>0$ is the transverse field and $J_q$ is the nearest-neighbor
	exchange. The transition separates a spin-ordered phase
	($g<|J_q|$), which spontaneously breaks the global $\mathbb Z_2$
	spin-flip symmetry, from a paramagnet ($g>|J_q|$) \cite{pfeuty}.
	Under the Jordan--Wigner transformation, spin flip becomes fermion
	parity and the chain takes the form of a Kitaev chain. Its
	$\mathbb Z_2$ invariant is the sign of the product of the
	fermion masses at momenta $0$ and $\pi$,
	$\nu_{\rm M}=\operatorname{sgn}[(g-J_q)(g+J_q)]$ \cite{kitaev}.
	At $g=|J_q|$, one mass crosses zero, closing the bulk gap and
	changing the invariant. For $g<|J_q|$, the invariant is $-1$ and
	an open chain supports a Majorana mode at each end in the
	long-chain limit. For $g>|J_q|$, the invariant is $+1$.
	The Ising order parameter and the fermion-parity invariant
	therefore describe the same quantum critical point in the spin
	and fermion variables, respectively.

\section{Local weights and regular-point Hamiltonian}

Let $x=\tanh K_1$, $y=\tanh K_2$, and $z=\tanh K_3$. We work near the
strong-bond boundary $x=1$ and use $|y|,|z|<1$. The limit $x=1$ means $K_1\to+\infty$.
We use the two-spin basis $(00,01,10,11)$, with
$Z|0\rangle=|0\rangle$ and $Z|1\rangle=-|1\rangle$.
Let $R_s(x,y,z)$ denote the polynomial sign-alternating vertex matrix
obtained by removing the overall scalar in Eq.~(4.29) of Ref.~\cite{kag}.
At the reference point $(x,y,z)=(1,0,0)$ it equals
$P_s=P(Z\otimes\one)$, where $P|ab\rangle=|ba\rangle$ exchanges the
two factors, $X,Y,Z$ are Pauli matrices, and $\one$ is the identity
on the relevant space. We right-multiply by $P_s^{-1}$ and divide by
the scalar normalization $N$ to define
\begin{equation}
 Q=\frac{R_sP_s^{-1}}{N},\qquad
 N=\frac{(1+x)^2(1-y^2z^2)}{4}.
 \label{eq:normalization}
\end{equation}
In particular, $Q$ is normalized to have trace four. Direct multiplication
gives the  following Pauli matrix expansion
\begin{align}
 Q={}&\one+\kappa X\otimes X
 +d\biggl[\frac{1+yz}{1-yz}Z\otimes\one
          +\frac{1-yz}{1+yz}\one\otimes Z \notag\\
 &\qquad
 +i\frac{(y-z)^2}{1-y^2z^2}X\otimes Y
 -i\frac{(y+z)^2}{1-y^2z^2}Y\otimes X\biggr]
 +d^2\bigl(Z\otimes Z+\kappa Y\otimes Y\bigr),\notag\\
 d={}&\frac{1-x}{1+x},\qquad
 \kappa=\frac{y^2-z^2}{1-y^2z^2}.
 \label{eq:pauli}
\end{align}
\subsection{Yang--Baxter family and RTT relation}

The coefficient $d$ vanishes at the strong-bond boundary. The
remaining coefficient $\kappa$ measures the imbalance of the other
two bond directions. Set $\kappa=\tanh s$ and define
$\R_0(s)=\one+\tanh s\,X\otimes X$. This is the
\emph{strong-bond Yang--Baxter family}: varying the spectral
parameter $s$ changes the normalized local weights while the
strong bond remains at its boundary value. Since
$\R_0(s)=\exp(sX\otimes X)/\cosh s$, the braid equation follows
from $[X_1X_2,X_2X_3]=0$ and additivity of the exponential
arguments. A site index denotes the space on which the Pauli
matrix acts.

To make the algebraic Bethe ansatz structure \cite{Takhtajian-book} explicit, let
$R_0^{\rm ord}(s)=P\R_0(s)$. For a periodic row with $L$ sites, the
monodromy and transfer matrix are
\begin{equation}
 M_a(u)=R^{\rm ord}_{0,aL}(u)\cdots R^{\rm ord}_{0,a1}(u),
 \qquad T_0(u)=\operatorname{tr}_a M_a(u).
\end{equation}
The labels $a,b$ denote two-state auxiliary spaces. The
Yang--Baxter equation gives the RTT relation
\begin{equation}
 R^{\rm ord}_{0,ab}(u-v)M_a(u)M_b(v)
 =M_b(v)M_a(u)R^{\rm ord}_{0,ab}(u-v).
 \label{eq:rtt}
\end{equation}
Tracing over both auxiliary spaces gives $[T_0(u),T_0(v)]=0$.
At $u=0$, $\R_0(0)=\one$ and $R_0^{\rm ord}(0)=P$, so
$T_0(0)$ is the one-site translation. The first local generator
$H=-T_0'(0)T_0(0)^{-1}$ is

\begin{equation}
 H_{\rm YB}=-\sum_j X_jX_{j+1}.
 \label{eq:ybham}
\end{equation}
In this ordinary realization the transfer matrix can be diagonalized
directly. Let $\eta_j=\pm1$ be the eigenvalue of $X_j$ in a
product state and take $\eta_{L+1}=\eta_1$. The row translates
this configuration and multiplies it by
$\prod_{j=1}^{L}(1+\tanh u\,\eta_j\eta_{j+1})$.
 A cyclic shift permutes the bond factors in this product,
	so the scalar is the same for every configuration in the translation
	orbit. On that orbit, each $T_0(u)$ therefore acts as the same cyclic
	shift multiplied by a scalar that depends on $u$. Fourier sums over
	the orbit diagonalize the shift and are simultaneous eigenstates of
	$T_0(u)$ for all $u$.
	In the original kagom\'e vertex convention, the local matrix contains
	the graded permutation $P_s$. The anisotropic scaling below uses the
	row normalization associated with this convention.

\section{Exact intersection with the classical critical branches}

Let $p,q$ be the dimensionless momentum variables in the convention
of Ref.~\cite{kag}. The Bloch phases along the two unit-cell
translations are $e^{2ip}$ and $e^{2iq}$, so each momentum is defined
modulo $\pi$. The classical fermionic spectral polynomial
$\mathcal P(p,q)$ vanishes where the fermionic kernel has a zero
eigenvalue. In Eq.~(4.34) of Ref.~\cite{kag}, it is
\begin{align}
 \mathcal P(p,q)&=A+B\cos(2p)+C\cos(2q)+D\cos[2(p-q)],\\
 A&=(1+x^4)(1+y^4)(1+z^4)-8x^2y^2z^2,\\
 B&=2x^2(1-y^4)(1-z^4),\\
 C&=2z^2(1-x^4)(1-y^4),\\
 D&=2y^2(1-x^4)(1-z^4).
\end{align}
At the three nonzero momenta $\mathbf k_i=(p_i,q_i)$ listed below,
the polynomial is a square, $\mathcal P(\mathbf k_i)=m_i^2$.
We call $m_i$ its signed mass factor. The critical branch in this
channel is the set of couplings where $m_i=0$. Explicitly,
\begin{align}
 m_1&=(1-x^2)(1+y^2z^2)+(1+x^2)(y^2-z^2),
 &\mathbf k_1&=(\pi/2,\pi/2),\\
 m_2&=(1+x^2)(1-y^2z^2)-(1-x^2)(y^2+z^2),
 &\mathbf k_2&=(0,\pi/2),\\
 m_3&=(1-x^2)(1+y^2z^2)-(1+x^2)(y^2-z^2),
 &\mathbf k_3&=(\pi/2,0).
 \label{eq:masses}
\end{align}
At fixed $y,z$, let $x_c$ be the value of $x$ at which either
$m_1$ or $m_3$ vanishes. Solving the two equations gives

\begin{equation}
 x_c^2=\frac{1+y^2z^2-|y^2-z^2|}
             {1+y^2z^2+|y^2-z^2|}.
 \label{eq:critical}
\end{equation}
The branch is $\mathbf k_1$ when $z^2>y^2$ and $\mathbf k_3$ when
$y^2>z^2$. The second branch does not reach $x=1$ for $|y|,|z|<1$.

At $x=1$ the full momentum dependence simplifies to
\begin{equation}
 \mathcal P(p,q)=4(y^2-z^2)^2
 +2(1-y^4)(1-z^4)[1+\cos(2p)] .
 \label{eq:line}
\end{equation}
The exact intersection with the verified Yang--Baxter family is thus
\begin{equation}
 x=1,\qquad y^2=z^2,\qquad
 p=\pi/2\pmod{\pi},\quad q\ \text{arbitrary}.
\end{equation}
The condition $y^2=z^2$ gives $\kappa=0$, so the classical
intersection selects the regular RTT point $s=0$.  The line of
zeros arises because the classical polynomial loses its $q$
dependence there. The quantum Hamiltonian generated by
differentiation along the boundary has a finite excitation gap
above its ordered ground-state sector.

\subsection{Exact fermionic decay length}
\label{sec:decay}

We now allow $q$ to be complex and fix $p=\pi/2$, the momentum
channel containing both branches that reach the boundary. The
imaginary part of the Bloch angle $2q$ gives the inverse decay
length $\gamma=\xi_{\rm f}^{-1}$ in units of the corresponding
lattice cell.  This use of complex momentum follows
	the general relation between the distance of a Fourier singularity
	from the real axis and the exponential decay rate \cite{Kohn1959}.
	Here, a Fourier factor $e^{2iqn}$ has magnitude
	$e^{-n\,\operatorname{Im}(2q)}$ for $n>0$. Let $U=(1-x^2)(1+y^2z^2)$ and
$V=(1+x^2)(y^2-z^2)$. Then $m_1=U+V$ and $m_3=U-V$.
At fixed $p=\pi/2$ the polynomial is linear in $\cos(2q)$.
Its values at $q=0$ and $q=\pi/2$ determine it completely:

\begin{equation}
 \mathcal P(\pi/2,q)=U^2+V^2-2UV\cos(2q)
 =(U-Ve^{2iq})(U-Ve^{-2iq}).
\end{equation}
For \(U>0\) and \(V\ne0\), the two roots in \(e^{2iq}\) are
\(U/V\) and \(V/U\). Equivalently, \(2q=2q_0\pm i\gamma\)
modulo \(2\pi\), with \(q_0=0\) for \(V>0\) and
\(q_0=\pi/2\) for \(V<0\). The inverse fermionic decay length
in the direction conjugate to \(2q\), measured per lattice cell, is
\begin{equation}
\displaystyle
 \gamma=\xi_{\rm f}^{-1}
 =\left|\ln\frac{U}{|V|}\right|
 =\left|\ln\frac{(1-x^2)(1+y^2z^2)}
                   {(1+x^2)|y^2-z^2|}\right| .
 \label{eq:exact-decay}
\end{equation}
Indeed, the root inside the unit circle gives an envelope
\(e^{-\gamma |n|}\) for a fermionic propagator at this fixed
transverse momentum with nonzero residue, where \(n\) counts
cells in the decay direction. For example, the Fourier
coefficient of the reciprocal polynomial is
\begin{equation}
 \int_{-\pi/2}^{\pi/2}\frac{dq}{\pi}\,
 \frac{e^{2iqn}}{\mathcal P(\pi/2,q)}
 =\frac{e^{2iq_0 n}e^{-\gamma|n|}}{|U^2-V^2|},
 \qquad U\ne |V|.
\end{equation}
Near either branch, away from the point where $U=V=0$, \(\gamma\) vanishes as
\(\lvert U-|V|\rvert/|V|\), and \(\xi_{\rm f}\) diverges
inversely with the distance to the branch along a transverse
path. When exactly one of \(U,V\) vanishes, the polynomial
is a nonzero constant in \(q\). The joint zero \(U=V=0\)
has a full line of zeros, as in Eq.~\eqref{eq:line}.
The length in Eq.~\eqref{eq:exact-decay} is defined in the
fermionic channel at fixed \(p=\pi/2\). 
%{\color{red}Extracting a spin--spin
%correlation length requires the corresponding spin correlator.?}

\section{Simultaneous anisotropic scaling}

Let $t>0$ be a dimensionless scaling parameter that tends to zero.
We keep $h>0$ and the real numbers $a,b$ fixed, with
$e^{-K_1}=ht$, $K_2=at$, and $K_3=bt$.  These transformations connect the classical 2D system with the 1D quantum problem \cite{kag,Baxter-book}.  The effective quantum Hamiltonian $H_{\rm aniso}$ is
	the negative coefficient of $t^2$ in the expansion of the normalized
	row transfer matrix $T(t)T(0)^{-1}$. With $\Delta\tau=t^2$ interpreted
	as a step in imaginary time, this normalized matrix agrees with
	$e^{-\Delta\tau H_{\rm aniso}}$ through order $t^2$.
Then $d=h^2t^2$ exactly and
$\kappa=(a^2-b^2)t^2+O(t^4)$. Equation~\eqref{eq:pauli} yields
\begin{equation}
 Q=\one+t^2\bigl[
 h^2(Z\otimes\one+\one\otimes Z)
 +(a^2-b^2)X\otimes X\bigr]+O(t^4).
 \label{eq:cardy}
\end{equation}
In the basis above all four off-diagonal entries at order $t^2$ have
the same coefficient. Thus the mixed $X\otimes Y$ and $Y\otimes X$
terms of the full normalized matrix enter only at order $t^4$ in this
scaling. 
%{\color{red}This expansion resolves an ambiguity in the displayed matrix
%of Eq.~(4.32) of Ref.~\cite{kag} and reproduces the couplings of
%Eq.~(4.33).?}

Let $T(t)$ be the row built from the original local factors
$R_s/N=QP_s$ with an ordinary auxiliary trace. The unperturbed
row $T(0)$ is the lattice shift times $\prod_j Z_j$. Normalize
the row by forming $T(t)T(0)^{-1}$, which tends to the identity.
At order $t^2$ one inserts the two-site term of
Eq.~\eqref{eq:cardy} at each position in the row. The next
decorated swap carries the outgoing auxiliary factor to the
neighboring site and conjugates it by $Z$. This changes the sign
of the exchange and preserves the field. On an even periodic
chain a rotation by $\pi$ about $Z$ on alternating sites removes
the exchange sign. In this convention the generator is

\begin{equation}
 H_{\rm aniso}=-\sum_j
 \bigl[2h^2 Z_j+(a^2-b^2)X_jX_{j+1}\bigr].
\end{equation}
Let $g=2h^2$ be the transverse field and $J_q=a^2-b^2$ the exchange
coefficient of the quantum chain. The dimensionless chain momentum
$k$ is defined modulo $2\pi$, and $E(k)$ is the positive
single-fermion excitation energy. The Jordan--Wigner dispersion is
$E(k)=2\sqrt{g^2+J_q^2-2gJ_q\cos k}$ \cite{pfeuty}.
For $J_q\ne0$ the gap closes at $g=|J_q|$. Independently, the
classical masses in Eq.~\eqref{eq:masses} have the expansions
\begin{equation}
 m_{1,3}=2t^2\bigl[2h^2\pm(a^2-b^2)\bigr]+O(t^4),
 \qquad m_2=2+O(t^2).
\end{equation}
For $J_q=a^2-b^2>0$, the classical factor $m_3$ becomes
$2t^2(g-J_q)$ to leading order, and the quantum gap closes at
$k=0$. For $J_q<0$, the factor $m_1$ becomes $2t^2(g+J_q)$
and the gap closes at $k=\pi$. These are the two possible
gap closings of the anisotropic generator. The regular RTT
derivative varies $\kappa$ at $d=0$ and gives the
pure-exchange Hamiltonian of Eq.~\eqref{eq:ybham}.

\begin{figure}[tbp]
 \centering
 \includegraphics[width=\linewidth]{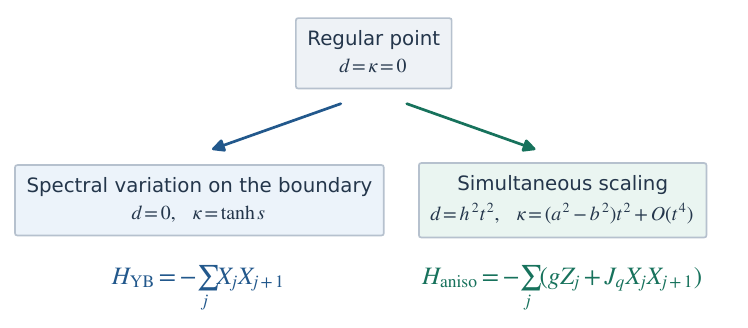}
 \caption{Two transfer-matrix constructions near the regular point.
 The spectral derivative on the strong-bond boundary stays at $d=0$.
 In the simultaneous scaling, $d$ and $\kappa$ enter at the same
 order in $t$. The diagram shows the limiting constructions
 schematically.}
 \label{fig:two-limits}
\end{figure}

Figure~\ref{fig:two-limits} shows which parameters vary in each
construction and which quantum Hamiltonian results.

\subsection{Fermionic topology of the anisotropic quantum chain}
\label{sec:topology}

Here $L$ is the number of sites, and an empty product in a
Jordan--Wigner string  \cite{pfeuty} equals the identity. The Majorana operators
$\alpha_j,\beta_j$ are Hermitian and satisfy
$\{\alpha_i,\alpha_j\}=\{\beta_i,\beta_j\}=2\delta_{ij}\one$ and
$\{\alpha_i,\beta_j\}=0$.

On an open chain define the Jordan--Wigner Majorana operators
\begin{equation}
 \alpha_j=\left(\prod_{\ell<j}Z_\ell\right)X_j,\qquad
 \beta_j=\left(\prod_{\ell<j}Z_\ell\right)Y_j.
\end{equation}
 The factors of $i$ follow from
	$\alpha_j\beta_j=iZ_j$ and
	$\beta_j\alpha_{j+1}=iX_jX_{j+1}$.
With these conventions the anisotropic Hamiltonian becomes
\begin{equation}
 H_{\rm aniso}
 ={\color{blue}i} g\sum_{j=1}^{L}\alpha_j\beta_j
  +{\color{blue}i} J_q\sum_{j=1}^{L-1}\beta_j\alpha_{j+1},
 \qquad g=2h^2,\quad J_q=a^2-b^2 .
\end{equation}
This is the Kitaev chain with equal magnitudes of hopping and
pairing. In the original spin variables, fermion parity acts as the global spin flip
$\prod_j Z_j$. For a gapped bulk , the parity-protected
$\mathbb Z_2$ invariant is the sign of the product of the two
Majorana masses at the particle--hole invariant momenta
$k=0,\pi$ \cite{kitaev}:
\begin{equation}
 \nu_{\rm M}=\operatorname{sgn}[(g-J_q)(g+J_q)]
 =\operatorname{sgn}\!\left[4h^4-(a^2-b^2)^2\right].
\end{equation}
When exchange dominates ($g<|J_q|$), the invariant is $-1$ and the fermionic chain is topological. When the field dominates ($g>|J_q|$), the invariant is $+1$ and the fermionic chain is trivial.

The operators $\Gamma_{\rm L}$ and $\Gamma_{\rm R}$ below are
localized near the left and right boundaries. Their overall factors
can be chosen so that each squares to the identity. 
The length $\xi_{\rm M}$ is the exponential localization length of
their amplitudes, measured in chain sites.

For $g<|J_q|$, let $r=g/J_q$. The left and right Majorana
end modes have profiles
\begin{equation}
 \Gamma_{\rm L}\propto\sum_{j=1}^{L}r^{j-1}\alpha_j,\qquad
 \Gamma_{\rm R}\propto\sum_{j=1}^{L}r^{L-j}\beta_j .
\end{equation}
These are exact zero modes at their respective semi-infinite
boundaries. On a finite chain the uncancelled terms in the
commutators with $H_{\rm aniso}$ are proportional to $r^L$,
which gives an exponentially small parity splitting.
The localization length is $\xi_{\rm M}^{-1}=\ln(|J_q|/g)$
in lattice units for $0<g<|J_q|$.

The link to the   defined anisotropic sign-alternating classical kagom\'e model follows from
Eq.~\eqref{eq:masses}. In the anisotropic scaling,
$m_3/(2t^2)\to g-J_q$ and $m_1/(2t^2)\to g+J_q$.
 For a gapped quantum chain, the product of these
	limiting masses is nonzero. Its sign gives the fermion-parity
	invariant,
	\[
	\nu_{\rm M}=\operatorname{sgn}\!\left[
	\lim_{t\to0}\frac{m_3m_1}{4t^4}\right]
	=\operatorname{sgn}(g^2-J_q^2).
	\]
	At the gap closing, the limiting product vanishes and the
	gapped-phase invariant is undefined. The branch at $\mathbf k_3$
changes the invariant for $J_q>0$, and the branch at
$\mathbf k_1$ changes it for $J_q<0$. In the original spin
variables, the topological fermion regime is the ordered
Ising phase with spontaneous breaking of the spin-flip
symmetry. The invariant and the Majorana modes concern the
one-dimensional fermionic transfer-matrix Hamiltonian in
the anisotropic limit.

In the same anisotropic scaling,
\(U/(2t^2)\to g\) and \(V/(2t^2)\to J_q\). For
\(0<g<|J_q|\), Eq.~\eqref{eq:exact-decay} therefore gives
\(\gamma\to\ln(|J_q|/g)=\xi_{\rm M}^{-1}\).
Thus the complex-momentum decay rate of the classical
polynomial in this channel reduces to the Majorana end-mode
localization rate of the quantum chain.

\begin{figure}[tbp]
 \centering
 \includegraphics[width=\linewidth]{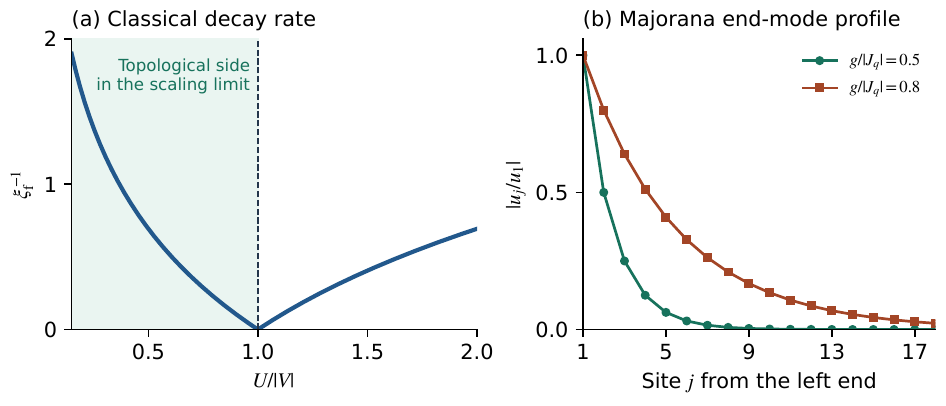}
 \caption{(a) Exact classical inverse fermionic decay length at
 $p=\pi/2$, from Eq.~\eqref{eq:exact-decay}. It vanishes at $U=|V|$.
 The shaded region maps to the topological side in the anisotropic
 limit. (b)  Normalized magnitude $|u_j/u_1|=(g/|J_q|)^{j-1}$
 	of the left Majorana mode for two illustrative ratios $g/|J_q|$,
 	where $u_j$ is the coefficient of $\alpha_j$ in
 	$\Gamma_{\rm L}=\sum_{j=1}^L u_j\alpha_j$.
 The curve closer to the critical ratio 1 decays more slowly.}
 \label{fig:fermion-majorana-decay}
\end{figure}

Figure~\ref{fig:fermion-majorana-decay} shows the vanishing
of the inverse decay length at the classical critical ratio and
the broadening of the Majorana profile as the quantum critical
point is approached.

\subsection{Group cohomology and fermionic SPT classification}
\label{sec:spt-classification}

\paragraph{Unitary spin flip}
For the spin Hamiltonian the on-site unitary symmetry is
 $G_{\rm spin}=\mathbb Z_2=\{1,\mathsf s\}$.
	Here $1$ is the identity element and $\mathsf s$ is the nonidentity
	element, with $\mathsf s^2=1$. Its action on the spin chain is
	represented by ${\mathsf U}(\mathsf s)=\prod_{j=1}^L Z_j$, which sends
	$X_j\mapsto -X_j$ and leaves $Z_j$ invariant at every site.
	The symbol $\mathsf s$ denotes this symmetry operation, while
	$s$ in $\kappa=\tanh s$ denotes the spectral parameter. A gaped 
	symmetry-preserving
short-range-entangled spin phase in one spatial dimension
has an edge projective class in $H^2(G_{\rm spin},U(1))$
\cite{cgw}. The edge representative ${\mathsf V}(s)$ is defined up to a phase.
A normalized two-cocycle $\omega$ records the phases in the
multiplication law of edge representatives, with the identity
representative fixed to $\one$. For $\mathbb Z_2$ its only
potentially nontrivial value is $\omega(s,s)=e^{i\phi}$, so
${\mathsf V}(s)^2=e^{i\phi}\one$.
Rephasing the edge operator by
${\mathsf V}(s)\mapsto e^{-i\phi/2}{\mathsf V}(s)$ sets ${\mathsf V}(s)^2=\one$.
Hence
\begin{equation}
 H^2(\mathbb Z_2,U(1))=0.
\end{equation}
The spin paramagnet $g>|J_q|$ is adiabatically connected to
the polarized product state by taking $J_q\to0$. The
ordered region $g<|J_q|$ has two thermodynamic ground states
exchanged by $U(s)$ and lies outside the classification of
symmetry-preserving spin ground states.

\paragraph{Fermion parity and the Arf invariant.}
For the fermionic locality defined by the Jordan--Wigner
variables, the symmetry group with parity alone is
$G_f=\mathbb Z_2^F$ and the quotient
$G_f/\mathbb Z_2^F$ is trivial. The corresponding interacting
classification in $1+1$ dimensions is the character group
of spin bordism \cite{kttw},
\begin{equation}
 \mathrm{Hom}\bigl(\Omega^{\mathrm{Spin}}_2,U(1)\bigr)
 \cong\mathbb Z_2.
\end{equation}
Here $\Omega^{\mathrm{Spin}}_2$ is the bordism group of closed
two-dimensional spin manifolds, with addition given by disjoint
union. A character is a homomorphism to $U(1)$. The Arf invariant
takes values in $\mathbb Z_2$, and $n_{\rm M}=0,1$ labels the trivial
and nontrivial fermionic classes, respectively.

On a closed surface $\Sigma$ with spin structure $\eta$, the
nontrivial character is the topological partition function
$(-1)^{\mathrm{Arf}(\eta)}$. In the chain of
Sec.~\ref{sec:topology}, its exponent is
\begin{equation}
 n_{\rm M}=\frac{1-\nu_{\rm M}}{2}
 =\frac{1-\operatorname{sgn}[(g-J_q)(g+J_q)]}{2},
 \qquad g\ne |J_q|.
\end{equation}
Thus the Arf phase is realized for $g<|J_q|$. At the
representative point $g=0$, the bulk Majoranas are paired
on bonds and $\alpha_1,\beta_L$ are unpaired. Within the
ground-state doublet the total parity acts, up to an overall
sign, as $i\alpha_1\beta_L$. Each end factor is parity odd.
For two copies, a local bilinear of the two Majoranas at the
same end preserves total parity and removes the boundary
degeneracy. This boundary algebra exhibits the order-two
stacking law directly.

\paragraph{Antiunitary refinement.}
The real-coupling chain also admits complex conjugation
$\mathcal T$ in the $Z$ basis, with $\mathcal T^2=1$.
It acts by $\alpha_j\mapsto\alpha_j$ and
$\beta_j\mapsto-\beta_j$. With $\mathcal T$ imposed on
fermionic perturbations, the quadratic chain is in
class BDI. Define $q(k)=g-J_q e^{ik}$, the complex off-diagonal
Bloch function in the chiral basis. The winding $w$ counts how
many times this function encircles the origin over the chain
Brillouin zone. With this convention,
\begin{equation}
 w=\frac{1}{2\pi}\int_{-\pi}^{\pi}
       \partial_k\arg q(k)\,dk
 =\begin{cases}
 1,&g<|J_q|,\\
 0,&g>|J_q|.
 \end{cases}
\end{equation}
The integer winding reduces to $\mathbb Z_8$ under
interactions preserving both symmetries, in agreement with
$\Omega_2^{\mathrm{Pin}^-}\cong\mathbb Z_8$
\cite{fidkowski,kttw}. Two Majoranas of the same
$\mathcal T$ parity at one end admit a parity-even bilinear
that is odd under $\mathcal T$. The extra symmetry therefore
protects the pair against this boundary coupling. The single topological chain
represents a generator in the stated winding convention.
Forgetting $\mathcal T$ retains $w\bmod 2=n_{\rm M}$.
These classifications refer to the $1+1$-dimensional
quantum chain defined by the anisotropic transfer-matrix
limit. The role of the two-dimensional classical kagom\'e
model here is to supply the critical mass factors that
determine where the chain changes class.

\section{A finite-coupling obstruction with one vanishing bond}

For $z=0$, $Q$ is proportional to
\begin{equation}
 W(x,f)=
 \begin{pmatrix}
 1&0&0&xf\\
 0&x&x^2 f&0\\
 0&f&x&0\\
 xf&0&0&x^2
 \end{pmatrix},
 \qquad f=y^2.
 \label{eq:W}
\end{equation}
A colored embedding means that $x$ and $f$ in $W$ may depend
on the two spectral variables separately \cite{ks,kh}.
We assume an analytic family in the fixed basis of
Eq.~\eqref{eq:W}, satisfying the braid equation and
$W(u,u)=\one$. In this calculation only, $(a,f)$, $(b,g)$,
and $(c,h)$ denote the values of $(x,f)$ at $(u,v)$, $(u,w)$,
and $(v,w)$, respectively, where $u,v,w$ are spectral parameters.
The quantities $A,F,C,G$ are the first-order Taylor coefficients
of the first and third pairs as $v$ varies with $u,w$ fixed.
At $v=u$ the pairs are $(1,0),(b,g),(b,g)$. For
$v=u+\epsilon$ with small $\epsilon$, the expansion is
\[
 (a,f)=(1,0)+\epsilon(A,F)+O(\epsilon^2),\qquad
 (c,h)=(b,g)+\epsilon(C,G)+O(\epsilon^2).
\]
Form the $8\times8$ braid residual
\begin{multline}
 (W(a,f)\otimes\one)(\one\otimes W(b,g))(W(c,h)\otimes\one)\\
 -(\one\otimes W(c,h))(W(b,g)\otimes\one)
   (\one\otimes W(a,f)).
\end{multline}
In its first-order coefficient, take the entries $(1,4),(1,6),
(1,7),(2,2)$ with one-based indexing in the lexicographic three-spin
basis. Their $4\times4$ coefficient matrix in the column order
$(A,F,C,G)$ has determinant
\begin{equation}
 -2b^4g^2(b^2-1)(b^2+1)(g^2-1).
 \label{eq:minor}
\end{equation}
For $0<b,g<1$ the determinant is nonzero, so all four
first-order coefficients must vanish. In particular,
$(C,G)=0$ forces the derivative with respect to the first
spectral variable of $W(u,w)$ to vanish. On a connected
analytic parameter domain that contains the regular diagonal,
this would make the matrix independent of that variable.
Regularity then forces the constant value $\one$. The
obstruction therefore applies to a nonconstant regular
analytic family in the fixed basis of Eq.~\eqref{eq:W} on
$z=0$, including the classical critical points
$x^2=(1-y^2)/(1+y^2)$. Families that vary $z$, staggered
intertwiners and spectral changes of basis are outside this test.

\section{Conclusion}

The two classical critical branches that reach the strong-bond
boundary identify the topological transition of the quantum chain in
the simultaneous anisotropic scaling. The scaled signed mass factors
are $g-J_q$ and $g+J_q$. Their product determines the fermion-parity
invariant, and its sign changes when a single mass crosses zero.
The same classical polynomial gives an exact fermionic decay length
at $p=\pi/2$. On the topological side that length approaches the
Majorana end-mode localization length. This relates a bulk decay
scale of the classical lattice to a boundary observable of the
quantum chain.

On the boundary itself, the regular Yang--Baxter family yields
commuting row transfer matrices whose first generator is pure
Ising exchange. The classical line of zeros occurs at its regular
point. The differentiated braid equation gives a finite-coupling
obstruction when one bond direction is absent and the matrix
basis is fixed. Establishing a regular finite-coupling embedding
with all three bond directions present remains an open problem.

\section*{Funding}
This research was supported by the Armenian HESC through grant 24RL-1C024.

\end{document}